\pdfoutput=1
\documentclass[12pt]{article}
\usepackage[utf8]{inputenc}
\usepackage{tabularx}
    \newcolumntype{L}{>{\centering\arraybackslash}X}
\usepackage{graphicx}
\usepackage{float}
\usepackage{setspace}
\usepackage[utf8]{inputenc}
\usepackage{tabularx}
    \newcolumntype{L}{>{\centering\arraybackslash}X}
\usepackage{graphicx}
\usepackage{float}
\usepackage{setspace}
\usepackage{hyperref}
\usepackage{multirow}
\usepackage{makecell}
\usepackage[margin=1in]{geometry}
\usepackage[sorting=none,style=ieee]{biblatex}
\date{\vspace{-5ex}}

\usepackage{xr}

\newcommand*{\addFileDependency}[1]{
\typeout{(#1)}
\IfFileExists{#1}{}{\typeout{No file #1.}}
}\makeatother

\newcommand*{\myexternaldocument}[1]{%
\externaldocument{#1}%
\addFileDependency{#1.tex}%
\addFileDependency{#1.aux}%
}

\myexternaldocument{SI}

\begin{document}
\footnotesize
\title{Automated patent extraction powers generative modeling in focused chemical spaces}

\author{Akshay Subramanian\textsuperscript{1,*}, Kevin P. Greenman\textsuperscript{2,*}, Alexis Gervaix\textsuperscript{3},\\ Tzuhsiung Yang\textsuperscript{4}, Rafael G{\'o}mez-Bombarelli\textsuperscript{1}}   

\maketitle

\begin{flushleft}
\textbf{1} Department of Materials Science and Engineering, Massachusetts Institute of Technology, Cambridge, MA, USA
\\
\textbf{2} Department of Chemical Engineering, Massachussets Institute of Technology, Cambridge, MA, USA
\\
\textbf{3} Swiss Airtainer SA, Yverdons-les-Bains, Vaud, Switzerland
\\
\textbf{4} Department of Chemistry, National Tsing Hua University, Hsinchu City, Taiwan
\\
\textbf{*} Contributed equally to this work
\end{flushleft}

\maketitle
\begin{center}
\end{center}


\begin{abstract}
Deep generative models have emerged as an exciting avenue for inverse molecular design, with progress coming from the interplay between training algorithms and molecular representations. One of the key challenges in their applicability to materials science and chemistry has been the lack of access to sizeable training datasets with property labels. Published patents contain the first disclosure of new materials prior to their publication in journals, and are a vast source of scientific knowledge that has remained relatively untapped in the field of data-driven molecular design. Because patents are filed seeking to protect specific uses, molecules in patents can be considered to be weakly labeled into application classes.  Furthermore, patents published by the US Patent and Trademark Office (USPTO) are downloadable and have machine-readable text and molecular structures. In this work, we train domain-specific generative models using patent data sources by developing an automated pipeline to go from USPTO patent digital files to the generation of novel candidates with minimal human intervention. We test the approach on two in-class extracted datasets, one in organic electronics and another in tyrosine kinase inhibitors. We then evaluate the ability of generative models trained on these in-class datasets on two categories of tasks (distribution learning and property optimization), identify strengths and limitations, and suggest possible explanations and remedies that could be used to overcome these in practice.
\end{abstract}

\section{Introduction}
The efficient navigation of chemical space for the design of novel candidate molecules has long been of interest to chemists and materials scientists. With the rapid surge in interest for data-driven approaches, deep generative models have emerged as an exciting avenue for inverse molecular design. \cite{schwalbe2020generative,elton2019deep} Progress in this field has come from the interplay between training algorithms and molecular representations. Over the last few years, approaches have used autoregressive, latent variable and reinforcement learning (RL) algorithms to generate string \cite{segler2018generating,bjerrum2017molecular,gomez2018automatic,kusner2017grammar,olivecrona2017molecular}, and graph \cite{mercado2021graph,jin2018junction,simm2020reinforcement,flam2021mpgvae} representations of molecules. While fully unsupervised models can be trained on large unlabeled data (for instance the 100+ million known, individually synthesized molecules from PubChem), inverse molecular design requires some form of supervision to steer generation towards high-performance molecules at the extremes of the property distribution. \cite{kim2016pubchem} One of the key challenges in the applicability of such inverse design models to materials science and chemistry has been the lack of accessibility to sizeable labeled training datasets in these fields. \cite{wu2018moleculenet}
\\

Published patents are an important source of scientific knowledge since the discovery of new materials and molecular candidates are disclosed in patents, years before their publication in scientific journals. \cite{senger2015managing,ohms2021current} Patent authorities such as the United States Patent and Trademark Office (USPTO), European Patent Office (EPO), Japanese Patent Office (JPO), and World Intellectual Property Organization (WIPO) make published patents accessible through their web interfaces. In the past decade, there has been significant progress in extracting and collating information from these sources programmatically to create large databases of chemical compounds \cite{papadatos2016surechembl}, and reactions \cite{lowe2012extraction}. This large body of extracted knowledge has immense potential in feeding 'data hungry' deep learning models, but has remained relatively untapped in the field of molecular design.\\

Since patents are filed seeking protection within a given application, they are thematically labeled into domains. This makes it relatively simple to extract domain-specific molecular structures. Moreover, they are likely to be high-performance since they merited the investment of a patent application, which allows us to create domain-specific generative models by training exclusively on molecules known to belong to the desired class. Our hypothesis is that training generative models on these smaller, but more meaningful datasets can automatically steer generation towards in-class high-performance molecules.
\\

All post-2001 chemistry patents published by the USPTO contain ChemDraw CDX, MDL, and TIFF files of chemical structures, as required by the Complex Work Unit (CWU) Pilot Program.\cite{CWU} This makes chemical structures more accessible in a computer readable format for large scale mining and screening efforts. In our work, we attempt to bridge the gap between these bulk data sources and data-driven chemical design, by developing an automated pipeline to isolate chemical structures from USPTO patents based on relevance to user-defined keywords, and demonstrating their utility as training data for deep generative models for molecular design. We choose three model types JTVAE \cite{jin2018junction}, RNN+SELFIES \cite{polykovskiy2020molecular,hochreiter1997long}, and REINVENT+SELFIES \cite{olivecrona2017molecular} to explore a variety of representations (graph, SELFIES\cite{krenn2020self}, and SELFIES respectively) and training algorithms (latent variable, autoregressive, and RL respectively), and show their applicability to learn data distributions in two patent-mined datasets that explore very different areas of the chemical space, i.e., organic photodiodes (OPD) and tyrosine kinase inhibitors (TKI). 
\\

We then test the ability of these models to perform property optimization in each of the following cases: 1) the property being optimized can be predicted accurately and cheaply,  2) Oracle property predictor is expensive, so we only have access to a proxy neural network predictor trained on oracle property data. \cite{gao2022sample, aldeghi2022roughness, westermayr2023high} In the TKI case, we optimize for high structural similarity to held-out, FDA-approved TKI molecules. This is a means to test the ability of models to optimize a robust, well-defined objective function with a relatively narrow solution space. This is an example of case 1 since we can calculate the similarity between molecules cheaply without the need for an approximator. In the OPD case, we choose our optimization objective to be the identification of organic molecules with low optical gaps. This is an example of case 2 since we approximate expensive DFT-computed optical gaps with a neural network predictor. Materials with low optical gaps, especially those that are sensitive to wavelengths of light in the near infrared (NIR) region of the spectrum have seen a growing interest due to their ability to utilize a larger portion of the solar spectral range which was previously difficult to access. Their applications are diverse ranging from military equipment to biomedical and semi-transparent devices. \cite{xu2008direct,xu2017flexible,liu2017efficient,li2017high} 
\\

The key observations we make through our experiments are summarized as follows: 1) We identify that patent-mined datasets offer the ability to create focused in-domain datasets of high-performing molecular structures. Training generative models on these datasets allows us to create in-domain generators that can generate novel candidates that model property distributions of the training data well. This offers a way to bootstrap focused domains of chemical space with limited human intervention.
2) Property optimization towards the edges of the training data distribution can be effective if we have access to a cheap oracle predictor, but is challenging when proxy neural network approximators are used. Proxy predictors are brittle (have the tendency to be adversarially attacked in our RL experiments), and difficult to train accurately end-to-end (learning properties from compressed latent space in JTVAE is difficult).

\section{Methods}
\subsection{Pipeline overview}
\begin{figure}[H]
  \centering
  \includegraphics[width=0.6\textwidth]{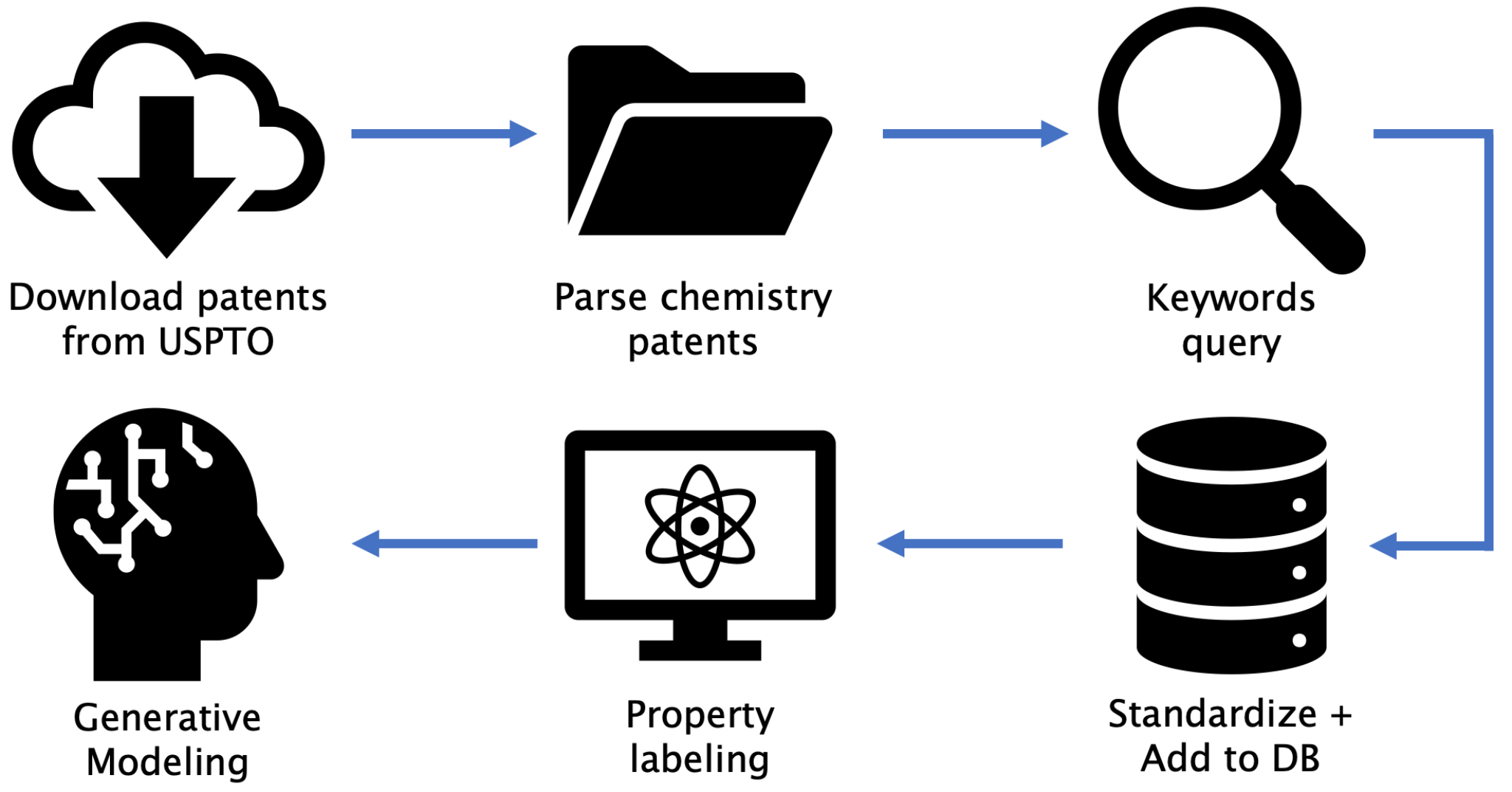}
  \caption{\textbf{Diagram of the workflow.} Patents are downloaded from USPTO, and chemistry patents are isolated. Keyword-based search is then performed to filter relevant patents and corresponding SMILES strings. A subset of molecules chosen based on computational budget are then labeled with properties. Generative models are trained to model the data distribution, which can be sampled to suggest novel candidates. }
  \label{fig:pipeline}
\end{figure}

Our overall pipeline consists of six steps: 1) Download patents from USPTO, 2) Parse chemistry patents, 3) Shortlist patents based on keywords, 4) Standardize data and add to our in-house database, 5) Property labeling for supervised property optimization tasks (DFT calculated optical gaps for OPD, and similarity to FDA-approved drugs for TKI), and 6) Generative modeling for distribution learning (unsupervised) and property optimization (semi-supervised). Figure \ref{fig:pipeline} shows a diagrammatic illustration of all steps involved. We make publicly available the code utilized in steps 1, 2, 3 and 6 along with this paper (URLs provided in Section \ref{data_availability}). Step 4 involved storage of all data in a database, followed by de-duplication of SMILES strings \cite{weininger1988smiles} and simple post-processing steps as described in Section \ref{SI_post_process} in the SI. A detailed description of procedures used in step 5 are provided in Section \ref{property_labeling}. These steps can be replaced by any form of data storage and property labeling technique depending on the chosen domain. An open source database framework similar to the one we used can be found at \cite{schwalbe2023mkite}.

\subsection{Patent extraction}
\label{extraction}
\begin{figure}[H]
  \centering
  \includegraphics[width=0.9\textwidth]{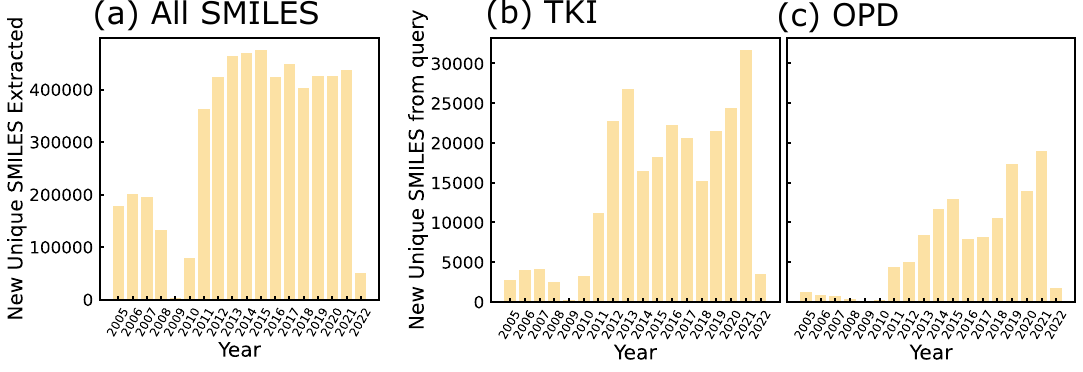}
  \caption{\textbf{Bar charts depicting number of SMILES strings extracted as a function of publishing year.} Strings extracted from patents published between 2005 and 2022 (a) before keyword-based filtering, (b) after application of TKI-based keyword search and (c) after application of OPD-based keyword search. SMILES were de-duplicated after sanitization by RDKit, such that all molecules within a given year are unique, and any molecule counted in a given year will not be counted in any future years. Years 2001-2004 are not shown and years 2008-2010 are incomplete due to inconsistencies in patent formatting (see Section 
  \ref{SI_patents} for details).}
  \label{fig:patent_composite}
\end{figure}

All granted USPTO patents from 2001 onward are available for download in weekly archives from the agency's Bulk Data Storage System (BDSS) at \url{https://bulkdata.uspto.gov/data/patent/grant/redbook/<YEAR>/}. We downloaded all of these archives from the BDSS using Python scripts by March 1, 2022. The compressed file size of all downloads was approximately 1.83 TB, including between 30 and 200 GB for each individual year. Next, we filtered out all patents that did not contain molecular structures in the form of \texttt{CDX} files. 
We encountered some difficulties in this filtering step with a subset of patent years due to inconsistent formatting and directory structures in the USPTO data (please refer to Section \ref{SI_patents} for details). For the remaining chemistry-related patents, we used RDKit \cite{landrum2013rdkit} to convert \texttt{MOL} files to SMILES strings. The number of new, unique SMILES strings extracted per year using this method are shown in Figure \ref{fig:patent_composite}a. We queried all chemistry-related patents by searching for keywords in each \texttt{XML} file. The TKI molecules shown in Figure \ref{fig:patent_composite}b were found using the keywords "tyrosine kinase inhibitor", and the OPD molecules in Figure \ref{fig:patent_composite}c are the result of querying for "organic electronic", "photodiode", and "organic photovoltaic". Any Markush structures in the dataset were filled in with ethyl groups because the particular substituents for each core molecule are not stored in a structured format that could be accessed without natural language processing; this included 17\% of molecules from the OPD query and 11\% of molecules from the TKI query. Thus, we generated a list of domain-relevant SMILES strings related to each set of keywords. More details on post-processing/filtering applied to the data are provided in Section \ref{SI_post_process}.

\subsection{Property labeling}
\label{property_labeling}
\subsubsection{TD-DFT calculations of optical gaps for OPD}
Initial conformations were generated with the ETKDG approach as implemented in RDKit, with at least 1500 attempts, up to 20 unique conformers were retained, ranking by their MMFF94 energies\cite{riniker2015better}. These geometries were refined using semi-empirical tight-binding density functional theory (GFN2-xTB) \cite{Bannwarth2019} in ORCA \cite{Neese2020}. Next, geometry optimizations were done at the BP86\cite{Becke1988}-D3\cite{Grimme2011}/def2-SVP\cite{Weigend2005} level of theory on the lowest-energy xTB conformer. Finally, TD-DFT calculations were performed with the Tamm-Dancoff approximation (TDA) \cite{hirata1999time} at the $\omega$B97X-D3\cite{Chai2009}/def2-SVPD level of theory in ORCA version 4.2.1. Reported optical gaps are the lowest-energy (reddest) singlet vertical excitation energies from the TD-DFT calculations.

\subsubsection{Similarity calculation for TKI}
Each TKI molecule was labeled with its Tanimoto similarity to Erlotinib, a held-out FDA-approved inhibitor. The Tanimoto similarity was computed over Morgan fingerprints of size 2048 and radius 2. The implementation for similarity and fingerprinting were both obtained from RDKit. While Erlotinib is the primary running example showed in this work, we also labeled molecules with similarity to the other 26 held-out inhibitors for similar experiments involving them (for ex. see Figure \ref{fig:SI_REINVENT_scores}).

\subsection{Generative modeling}
\subsubsection{Evaluation tasks}
We prepared two datasets: 1) OPD - Organic Photodiodes and 2) TKI - Tyrosine Kinase Inhibitors, covering two different chemical spaces. Models trained on these datasets were evaluated on two categories of tasks: 1) Distribution Learning - The ability of models to learn the training data distribution, and 2) Property Optimization - The ability of models to generate in-domain molecules that are optimized for a property of interest. Good  performance on the latter task would require some or all of the generated samples to be superior in properties in comparison to the training data distribution.
\\

For distribution-learning tasks, we evaluated models on the GuacaMol distribution learning benchmark metrics: Validity, Uniqueness, Novelty, KL Divergence and Frechet ChemNet Distance. \cite{brown2019guacamol,preuer2018frechet}. We also visualized the ground-truth property distribution of the sampled data and compared it with that of the training data. A close match between the two is an indicator of success in learning the training data distribution. For property optimization, we performed a similar visualization. Here, a shift in distribution towards higher values of the objective function is an indicator of good performance. Finally, to test the value of domain-focused training on property optimization, we compared the patent-trained models against baseline models that were trained on the ZINC dataset \cite{irwin2005zinc} but optimized for OPD and TKI properties. It is considered good performance if the domain-trained models generate molecules with more optimal properties than the generic model trained on the ZINC dataset. This would suggest that the structural priors imposed on the models by training on the domain-specific patent datasets reflect in more optimal properties for that domain. More specifics on the task formulation for each dataset are given below.
\\

For OPD tasks, the patent-mined OPD molecules were used as the training dataset. The property of interest in the distribution learning tasks was the DFT-computed optical gaps of sampled molecules. Since our aim was to generate molecular candidates with low optical gaps, the negative of the optical gaps as predicted by a proxy neural network predictor was used as the objective function which was maximized in the property-optimization tasks.
\\

For TKI tasks, the patent-mined TKI dataset was used as the training dataset. The property of interest in the distribution learning tasks was the similarity between sampled molecules and Erlotinib, an FDA-approved inhibitor, to gauge the model's ability to optimize a robust, well-defined objective function with a relatively narrow solution space. This quantity was also used as the objective function which was maximized in the property-optimization tasks. In addition to the tasks described earlier in this section, an additional distribution learning task was introduced for this dataset. Molecules sampled from models trained on TKI and ZINC datasets, and 27 held-out FDA-approved TKI molecules were projected on a 2-dimensional space with Principal Component Analysis (PCA). Samples from TKI-trained models lying closer than the ZINC-trained samples to the held-out molecules, would indicate that the models have accurately learned information about molecular structure from the training dataset. It is a way to test the utility that training on domain-focused data (TKI-patents) has over training on publicly accessible large databases (ZINC) that have a similar chemical space (drug-like molecules) but are less-focused on the domain of interest. Morgan fingerprints of size 2048 and computed with radius 2 was the molecular representation used during PCA.

\subsubsection{Generative models}
\label{models}
We evaluated two categories of generative models, i.e. 1) text-based and 2) graph-based, on these tasks. RNN+SELFIES and REINVENT+SELFIES fall under the first category while JTVAE falls under the second. RNN+SELFIES was only used for distribution learning tasks, REINVENT+SELFIES was used for only property optimization tasks, and JTVAE was used for both. SELFIES was used as the string representation of choice to ensure validity of structures generated. \cite{krenn2020self}
We go over some of the implementation choices for each below.
\\

Recurrent Neural Networks (RNNs) have been shown to be simple but powerful text-based models for distribution modeling tasks in molecules \cite{flam2022language}. They are trained using an auto-regressive training strategy where the next token is predicted at every time-step. The implementation from the MOSES Benchmarking platform \cite{polykovskiy2020molecular} was used with some modifications pertaining to change in representation from SMILES to SELFIES. The trained RNN can be sampled by feeding a BOS (beginning of sentence) token, and sampling the probability distribution predicted by the model autoregressively. An LSTM \cite{hochreiter1997long} network with 3 hidden layers and dropout probability of 0.2 between layers was used, with a final linear layer to transform the LSTM output into the required output sequence size. All LSTM hidden layers and the final linear layer were of size 768, and a learning rate of 1e-3 was used for the Adam optimizer. \\

Junction Tree Variational Autoencoder (JTVAE) is a graph-based generative model that learns to sequentially decode graph substructures using Message Passing Neural Networks (MPNNs), and combine them to form complete molecular structures. \cite{jin2018junction} It maintains a vocabulary of substructures decomposed from the training data, that are used during the decoding step to ensure validity of generated molecules. The model is trained by training the encoder, decoder and property predictors end-to-end with a multi-task loss function. Once trained, the latent space can be either randomly sampled or optimized by utilizing gradients from the property predictors. In both cases, the sampled latent vectors are passed through the decoder to obtain molecular candidates.
A graph Message Passing Network (MPN) with 3 layers was used in the graph encoder, and a graph GRU \cite{cho2014properties} network with 20 layers was used in the tree encoder, to form a concatenated latent vector of size 56. A learning rate of 1e-3 that was set to decay exponentially during the course of training was used for the Adam optimizer \cite{kingma2014adam}. More details given in Section \ref{SI_JTVAE}.
\\

REINVENT is a policy based Reinforcement Learning (RL) approach that learns to generate molecular structures optimized with a chosen objective function. \cite{olivecrona2017molecular} Training is performed in two steps: 1) A Prior RNN is pre-trained on a language modeling task, i.e., learning to predict the next token of the sequence by maximizing the likelihood on the training dataset. 2) Then, an augmented likelihood function is defined to be the sum of the Prior likelihood and a score indicating the desirability of the sequence. The agent, which is initialized with the Prior RNN weights, is then fine-tuned to minimize the squared difference between the agent likelihood and the augmented likelihood on samples drawn from the Agent RNN. Sampling from the trained model is performed in identical fashion to RNN (described in previous paragraph).
We once again use SELFIES representations of molecules. The Agent RNN was composed of three GRU cells \cite{cho2014properties}, each of size 512, followed by a linear output layer. Pre-training and fine-tuning were carried out using an Adam optimizer with learning rates of 1e-4 and 5e-4 respectively. We retained the same architectural choices used by Olivecrona et al. since our task of similarity-based optimization is nearly identical to the similarity guided structure generation experiments described in their work.

\section{Results and discussions}
\subsection{Distribution learning}
Table \ref{GuacaMol} compares the scores of the RNN+SELFIES and JTVAE models on the GuacaMol distribution learning benchmarks. Both models were able to generate molecules with relatively high validity, uniqueness and KL Divergence scores. We however found that JTVAE is superior to RNN+SELFIES in Novelty scores, and both models perform relatively poorly on Frechet ChemNet Distance scores. These observations may both be characteristics of the training datasets that we use being smaller and more domain-focused than the larger and more diverse drug datasets that have been benchmarked on these metrics in the past.
\\

\begin{figure}[H]
  \centering
  \includegraphics[width=0.9\textwidth]{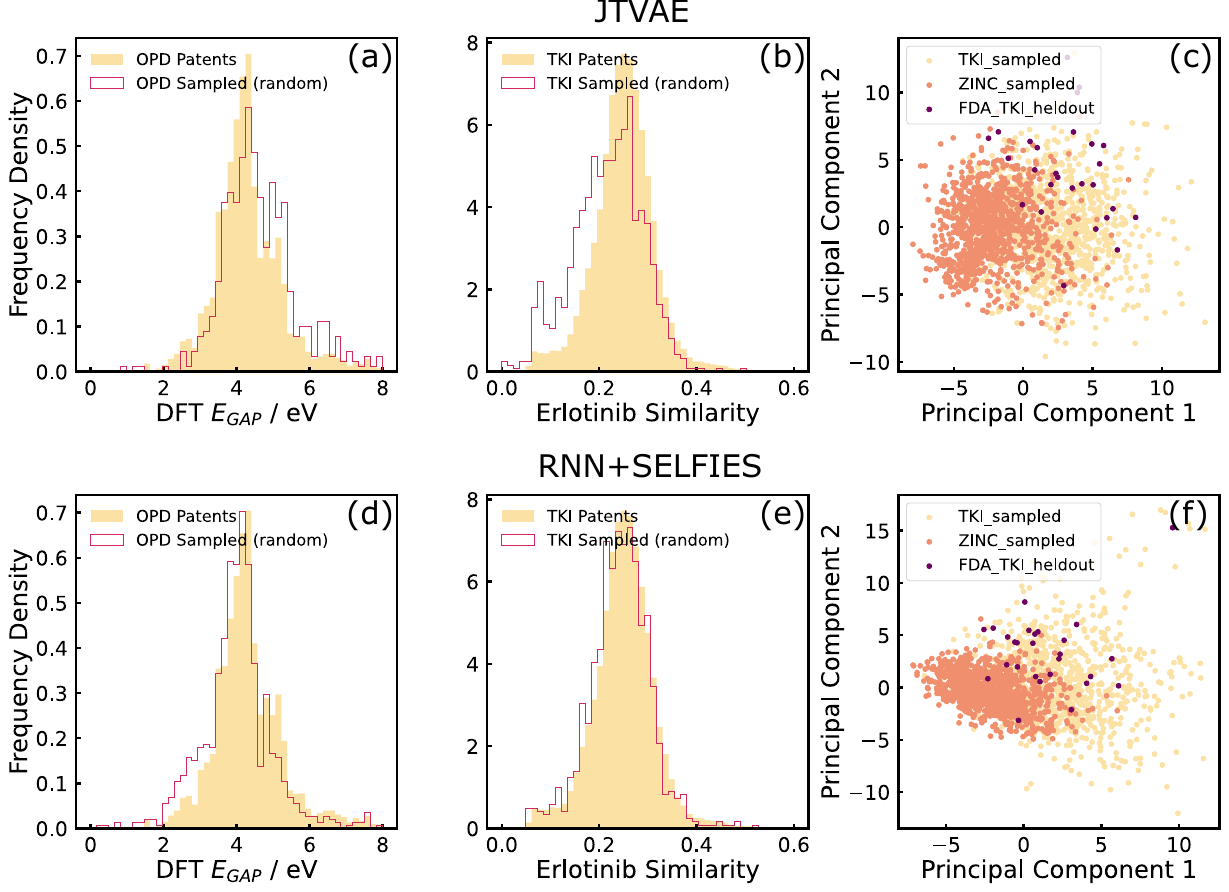}
  \caption{\textbf{Results on distribution learning tasks.} (a) and (b) show the property distributions of JTVAE-sampled molecules in comparison to training data properties, on OPD and TKI datasets respectively. (d) and (e) show the same distributions for molecules sampled from RNN+SELFIES. (c) and (f) show PCA projections of molecules randomly sampled from TKI-trained and ZINC-trained models, and held-out FDA approved inhibitors, for JTVAE and RNN+SELFIES respectively.}
  \label{fig:DL_composite}
\end{figure}

As can be seen in sub-figures (a), (b), (d), and (e) of Figure \ref{fig:DL_composite}, both models generated molecules whose properties matched well with the training dataset. It can also be observed that RNN+SELFIES is able to match the distributions better than the JTVAE, which conforms with the observations made by \cite{flam2022language}. Additionally, sub-figures (c) and (f) show that samples from TKI-trained models lie closer to held-out FDA-approved inhibitors than ZINC-trained samples, which indicates that both models have been able to learn structural information from the training datasets.
\\

From these results, we conclude that the deep generative models explored in this work are effective tools to model property distributions of arbitrary small, chemically focused, training datasets automatically extracted from the patent literature. The models can thus sample novel, in-distribution molecular structures that resemble the training data in terms of structure and properties. Furthermore, this suggests that domain-specific, focused chemical spaces can be boostrapped automatically from the literature without user-defined heuristics for the domain, as evidenced by the GuacaMol distribution learning benchmarks in two very distinct chemical spaces.

\begin{table}[htbp]
\centering
\caption{\textbf{GuacaMol distribution learning benchmarks for 1000 samples drawn from RNN+SELFIES and JTVAE, on OPD and TKI datasets.} Closer to 1.0 indicates better performance.}
\begin{tabular}{|c|c|c|c|}
\hline
\multirow{2}{*}{Model} & \multirow{2}{*}{Metric} & \multicolumn{2}{c|}{Dataset} \\
\cline{3-4}
                       &        & TKI & OPD \\ \hline
\multirow{5}{*}{\makecell{RNN+SELFIES \\ (random sample)}} & Validity & 1.00 & 0.99\footnotemark{}\\
                     & Uniqueness & 0.99 & 0.99 \\ 
                     & Novelty & 0.55 & 0.58 \\ 
                     & KL Divergence & 0.98 & 0.96 \\
                     & Frechet ChemNet Distance & 0.60 & 0.61 \\ \hline
\multirow{5}{*}{\makecell{JTVAE \\ (random sample)}} & Validity & 1.00 & 1.00 \\
                     & Uniqueness & 1.00 & 0.99 \\ 
                     & Novelty & 1.00 & 0.89 \\ 
                     & KL Divergence & 0.75 & 0.87 \\
                     & Frechet ChemNet Distance & 0.32 & 0.28 \\ \hline
\end{tabular}
\label{GuacaMol}
\end{table}
\footnotetext{ While rdkit does not process one generated molecule as valid, it is formally valid with bivalent lithium forming two covalent bonds.}

\subsection{Property optimization}
We evaluated generative models trained on patent-extracted, domain-focused datasets for property optimization. We evaluated REINVENT+SELFIES, which uses reinforcement learning and a string-based representation  and JTVAE, which performs optimization in the latent space and decodes locally optimal molecules, under this category of tasks. We identified that property optimization tasks towards edges of the training data distribution are challenging for a variety of reasons. We observed from our RL experiments that optimizers may push the designs out of the training domain which was particularly acute when a neural network predictor was used as a proxy for the oracle property. Here, the generative model can be thought of as performing an adversarial attack on the poorly-covered areas of the predictor. From our VAE experiments, we observed that it is sometimes challenging for proxy predictors to learn properties from compressed latent representations, and the unreliable objective function thereby leads to challenges in latent space optimization.
\\

Both these challenges arise from coupling generation and property optimization end-to-end. By instead splitting these into two separate steps of random sampling and post-hoc filtering, we observed better shifts in property histograms. More details on the post-hoc filtering approach are provided below in Section \ref{posthoc}.

\subsubsection{Post-hoc filter}
\label{posthoc}
We use the term "post-hoc filter" to refer to a property screen conducted on molecules that were randomly sampled from trained models. 
It can use either the predictions of a proxy predictor when the oracle property is expensive as in OPD tasks, or the oracle itself when it is cheap to compute as in TKI tasks. The degree of the filter applied (which we chose to be top 20\%) can be chosen based on the extent of screen to be performed. As a proxy predictor for OPD tasks, we trained a Chemprop MPNN model \cite{yang2019analyzing} on the patent-mined OPD dataset to predict DFT-calculated optical gaps (See \ref{fig:SI_chemprop_hexbin}). A random train-val-test split (60:20:20) was used to train, tune and evaluate the model.

\subsubsection{Approximate objective}
\label{approx}
\begin{figure}[H]
  \centering
  \includegraphics[width=0.9\textwidth]{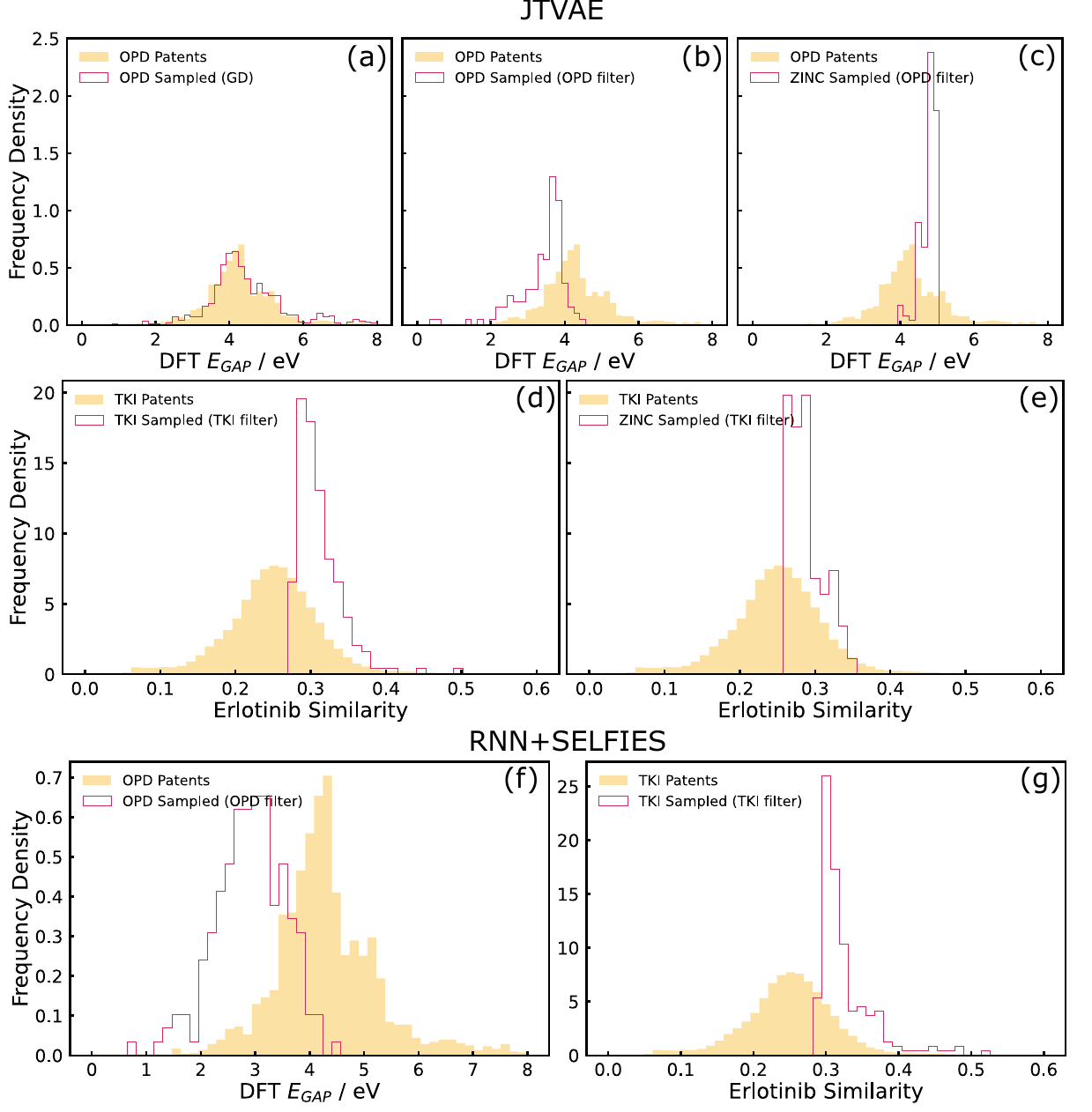}
  \caption{\textbf{Results on property optimization tasks.} (a), (b), (c), (d) and (e) show results for JTVAE, while (f) and (g) show results for RNN+SELFIES. (a) and (b) show OPD property distributions of molecules sampled by gradient descent (GD), and post-hoc filter respectively in comparison to training data properties. (c) shows the property distribution obtained by applying an OPD post-hoc filter to samples drawn from a ZINC-trained model. (d) and (e) are analogous to (b) and (c) but on TKI instead of OPD. (f) and (g) are analogous to (b) and (d) but on RNN+SELFIES instead of JTVAE.}
  \label{fig:PO_composite}
\end{figure}

All OPD optimization tasks required the use of a proxy neural network model since DFT simulations are computationally expensive and are typically not autodifferentiable, so it is not possible to train end-to-end generation and property scoring. \cite{ekström2010arbitrary, tamayo2018automatic} In the JTVAE case, a Multi Layer Perceptron (MLP) was used as a proxy predictor to predict oracle DFT-calculated optical gaps from the latent space. As can be seen from Figure \ref{fig:PO_composite} (a), gradient descent over the latent space in JTVAE has almost no effect in shifting property distributions away from the OPD training data towards lower optical gaps. To improve the optimization performance, we utilized the Chemprop posthoc filter to selectively isolate decoded candidates having predicted optical gaps below the the 20th percentile. This was useful in shifting the distribution towards lower optical gaps as can be seen from Figure \ref{fig:PO_composite}(b). The justification behind this approach was that learning properties from the latent space is a more challenging task than learning directly from the molecular graph. \cite{eckmann2022limo} The MLP predicting the optical gap from the latent space achieves an RMSE of 0.56 eV on the test set while the Chemprop model achieves an RMSE of 0.38 eV on the test set, which follows our intuition. The fact that JTVAE learns from a multi-task loss function composed of reconstruction and property terms, makes it a constrained optimization task that reduces the degrees of freedom of the MLP during training, and can hence make convergence more challenging. We observed similar challenges with coupling generators and property optimizers while training the REINVENT+SELFIES on the OPD dataset, where the Chemprop model described above was used as the proxy predictor modelling the reward function. Here, the generator could be thought of as performing an adversarial attack on the proxy predictor and converged at molecular candidates that optimized the proxy objective but were structurally unphysical. More details on JTVAE+MLP training are provided in Section \ref{SI_JTVAE} and details on REINVENT results on OPD data are provided in Section \ref{SI:REINVENT_OPD}.
\\

Apart from the described issue pertaining to the poor predictive performance of the MLP, there could be other potential reasons for the failure of gradient descent on the latent space.
One possibility is the presence of cascading effects. The unreliability of the MLP could have caused the points reached by gradient descent (on the latent space) to be outside the data distribution that the decoder saw during training, causing the decoder to be unreliable and collapse to a distribution more similar to the training data. One way to investigate this failure mode in the future could be the use of decoder uncertainty estimation techniques to identify such points and restrict samples to low-uncertainty regions of the decoder. \cite{notin2021improving} Another possibility is that the latent space manifold of the trained model was "rough" with respect to the MLP-predicted property, rendering optimization techniques such as gradient descent challenging. This could be investigated in more depth by evaluating the 'roughness' of the latent space with metrics such as roughness index (ROGI). \cite{aldeghi2022roughness} Therefore, it should be noted that the coupled interactions between generators and property predictors is a complicated problem, and utilizing approaches such as the post-hoc filter could be relatively simple remedies to these pitfalls even without a detailed knowledge of the failure mode. Demonstration of post-hoc filter with another model (RNN+SELFIES) is shown in Figure \ref{fig:PO_composite}(f), which can be used as a remedy for the adversarial attack issues observed in the REINVENT example again arising from coupling of generators (RNN) and property optimizers (RL).
\\

Finally, Figure \ref{fig:PO_composite}(c) is a baseline where training was performed on the ZINC dataset and post-hoc filters on the OPD target was applied. It can be clearly seen that sub-figure (b) is more shifted towards optimal properties than the ZINC baselines which suggests that the structural priors imposed by training on the domain-specific OPD patent dataset offers significant value in achieving optimal properties for that domain. For example, molecules incorporating structural priors such as conjugated rings have more potential in achieving low optical gaps than drug-like structures.
\\


\subsubsection{Oracle objective}
\label{oracle}
\begin{figure}[hbt]
  \centering
  \includegraphics[width=0.9\textwidth]{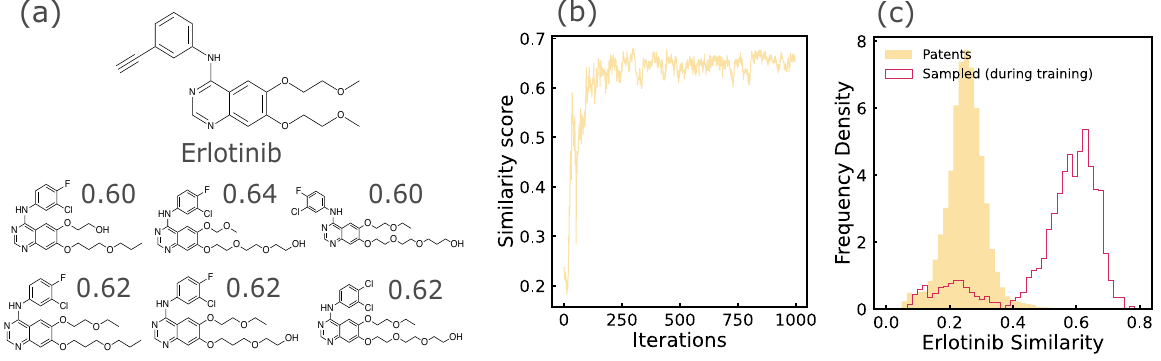}
  \caption{\textbf{Results based on REINVENT+SELFIES model trained on the TKI dataset.} (a) Candidates generated by REINVENT towards the end of training, with structural similarity to Erlotinib being the reward function. Similarity scores are indicated below each candidate. (b) Tanimoto similarity score computed between generated candidates and Erlotinib, as a function of training iterations. (c) Histograms showing properties of candidates sampled during agent training, in comparison with the training data distribution. }
  \label{fig:REINVENT_composite}
\end{figure}

In TKI optimization tasks, the property of interest was similarity to a chosen query structure, which is a cheap and oracle property estimate that can be calculated at every step of optimization.
In such cases where we have access to the oracle predictor, we observed better performance on optimization tasks. Figure \ref{fig:REINVENT_composite} (c) shows the Erlotinib similarity distribution of samples generated during training of REINVENT+SELFIES, which are clearly shifted towards higher values than the training data. (a) shows sample candidates along with their similarity scores, and (b) shows the improvement of similarity score as training progresses.
\\

A post-hoc filter using the oracle predictor can also be utilized in this case as a way to generate a set of novel candidate molecules that are optimized in comparison to the training data (Figure \ref{fig:PO_composite}(d) and (g) for JTVAE and RNN+SELFIES respectively). Similar to the example described in Section \ref{approx}, we also compared with a ZINC-trained baseline optimized for the TKI target, and observe minor improvements in shifts for the TKI-trained model in comparison to the ZINC-trained baseline (see Figure \ref{fig:PO_composite}(d) and (e)). This difference is not as significant as the OPD-ZINC baseline since the chemical spaces of ZINC and TKI datasets are fairly similar structurally.

\subsubsection{An alternative interpretation}
The above observations from JTVAE and REINVENT+SELFIES can also be interpreted with reference to terminology introduced by \cite{kajino2022biases}. While Kajino et al. primarily examine the existence of biases in Reinforcement Learning settings, the terminology can conceptually be extended to other types of generative models as well. In our tasks, both generative model and property predictor were trained on the same patent-mined dataset. This could have introduced reusing bias, which stems from effectively training and evaluating our model with information drawn from the same data source. In addition, during property optimization, the property predictor often sees unrealistic/nonphysical molecules which are far away from its training data distribution. This results in a misspecification bias, caused by the unreliability of the property predictor at points far away from the training data distribution. These two components of bias might have had a role to play in the observations we made in cases where a proxy predictor was used. Oracle property models on the other hand, are free from these two forms of bias.

\section{Conclusions}
In this work, we developed a framework to automatically extract molecular structures from the USPTO patent repository based on user-defined keyword searches, and generate datasets for machine learning in chemistry. We demonstrate the utility of the extracted datasets in training generative models for inverse molecular design tasks. We show that these datasets can be utilized to generate novel molecular structures with properties similar to the training dataset, in a completely unsupervised setting. We also evaluate model performance on supervised property optimization tasks, identify some limitations of existing models in shifting property distributions away from the training data regime, and suggest some possible explanations and remedies that could be used to overcome these in practice. The key observations we make through our experiments are summarized as follows: 1) We identify that patent-mined datasets offer the ability to create focused in-domain datasets of high-performing molecular structures and offers a way to bootstrap focused domains of chemical space with limited human intervention.
2) Property optimization towards the edges of the training data distribution can be effective if we have access to a cheap oracle predictor, but is challenging when proxy neural network approximators are used.

\section{Data and code availability}
\label{data_availability}
The code used to train models is publicly available. JTVAE: \url{https://github.com/wengong-jin/icml18-jtnn}, REINVENT: \url{https://github.com/MarcusOlivecrona/REINVENT}. The RNN models were trained using the char-rnn code from \url{https://github.com/molecularsets/moses}. A static version of the exact forks used is available at \url{https://doi.org/10.5281/zenodo.7719958}, and checkpoints of trained models and all training data including DFT-calculated properties are available at \url{https://doi.org/10.5281/zenodo.7996464}. \cite{akshay_subramanian_2023_7719959} Code for the patent mining and filtering pipeline can be found at \url{https://github.com/learningmatter-mit/PatentChem}. This patent code is archived at \url{https://doi.org/10.5281/zenodo.7719675}. \cite{greenman_kevin_2023_7719676} GuacaMol benchmarking was performed using \url{https://github.com/BenevolentAI/guacamol}

\section{Author contributions}
A.S. trained the generative models and analyzed the distribution learning and property optimization results. K.P.G. updated and organized the patent code and ran the high-throughput physics-based calculation pipeline. A.G. wrote an initial version of the patent code. T.Y. trained initial versions of the generative models. A.S. and K.P.G. wrote the first manuscript draft. R.G.-B. conceived the project, supervised the research, and edited the manuscript.

\section{Conflicts of interest}
There are no conflicts to declare.

\section{Acknowledgements}
A.S. was supported by funding from Sumitomo Chemical. K. P. G. was supported by the National Science Foundation Graduate Research Fellowship Program under Grant No. 1745302. This work was also supported by the DARPA Accelerated Molecular Discovery (AMD) program under contract HR00111920025. We acknowledge the MIT Engaging cluster and MIT Lincoln Laboratory Supercloud cluster \cite{reuther2018interactive} at the Massachusetts Green High Performance Computing Center (MGHPCC) for providing high-performance computing resources to run our TD-DFT calculations and train our deep learning models. 

\printbibliography[heading=bibnumbered]

\renewcommand\thesection{S\arabic{section}}
\setcounter{section}{0}
\renewcommand\thefigure{S\arabic{figure}}
\setcounter{figure}{0}
\renewcommand\thetable{S\arabic{table}}
\setcounter{table}{0}

\section{Patent Format Inconsistency}
\label{SI_patents}
As described in Section 2.2, the USPTO makes available machine-readable patents from 2001 to the present. However, these files are not consistent in their format and directory structure. As a result of these inconsistencies, our original extraction pipeline omitted years 2001-2004 because these years used SGML 2.4 or XML 2.5, whereas years 2005-present used XML 4.0-4.7, as described at \url{https://bulkdata.uspto.gov/}. Additionally, patents from late 2008 to early 2010 were omitted by our original pipeline because of a different directory structure than other patent releases. The initial training dataset for our generative models omitted some or all patents from the aforementioned years. Since our goal is to incorporate structural priors from a general region of domain-relevant chemical space rather than to extract a comprehensive set of domain-relevant molecules, this omission does not invalidate the approach. As we demonstrate, our approach is helpful for focusing chemical space even while omitting all patent years prior to 2001 (since they are not machine readable). For the same reason, our approach still works while omitting a subset of years after 2001. However, for the sake of completion and to maximize training dataset coverage of relevant structures, we have resolved these issues  in the latest version of our PatentChem code (\url{https://github.com/learningmatter-mit/PatentChem}). Going forward, users who do their own keyword queries with our code will be unaffected by the problems we initially encountered with certain years. 

\section{Processing of patent-extracted data before model training}
\label{SI_post_process}
The goal of our pipeline is to generate structures with limited domain knowledge beyond keywords, so we kept processing/filtering to a minimum except for constraints that allowed for better computational tractability and basic filters on molecular mass. For example, we applied a 1000 g/mol  maximum molecular mass cutoff on the OPD dataset primarily because JT-VAE has a sequential decoding process that enumerates combinations of fragment pairs, which scales with the size of fragments and is thus very slow for large molecules. This has the added benefit of eliminating polymers and large candidates (non-ideal for deposition techniques such as chemical vapor deposition). Similarly on the TKI dataset, we imposed maximum and minimum cutoffs of 700 g/mol and 250 g/mol respectively to eliminate candidates that are not "drug-like". We apply the minimum molecular mass constraint in the TKI case since our property optimization objective was similarity to held-out FDA approved drugs whose molecular masses typically fall above 250g/mol. 
\\

Our minimal filtering means there are some structures in our training datasets that are not domain-relevant (such as reagents or intermediates). However, the “false positives” (molecules that the model generates because it thinks they are relevant, when in reality they are not relevant) that come from this can be easily filtered out by the property labeling step. Just as a user can choose their own property-labeling method appropriate for their design task when using our code, they could also insert additional domain-knowledge-based preprocessing of the training dataset. Our current work demonstrates that the approach can still be useful even without this preprocessing, but additional filtering may improve results in some domains. We have provided some options for possible filters in our PatentChem code, such as minimum and maximum molecular weight and charged/neutral molecules.

\section{REINVENT+SELFIES}
\subsection{TKI}
\begin{figure}[H]
  \centering
  \includegraphics[width=0.9\textwidth]{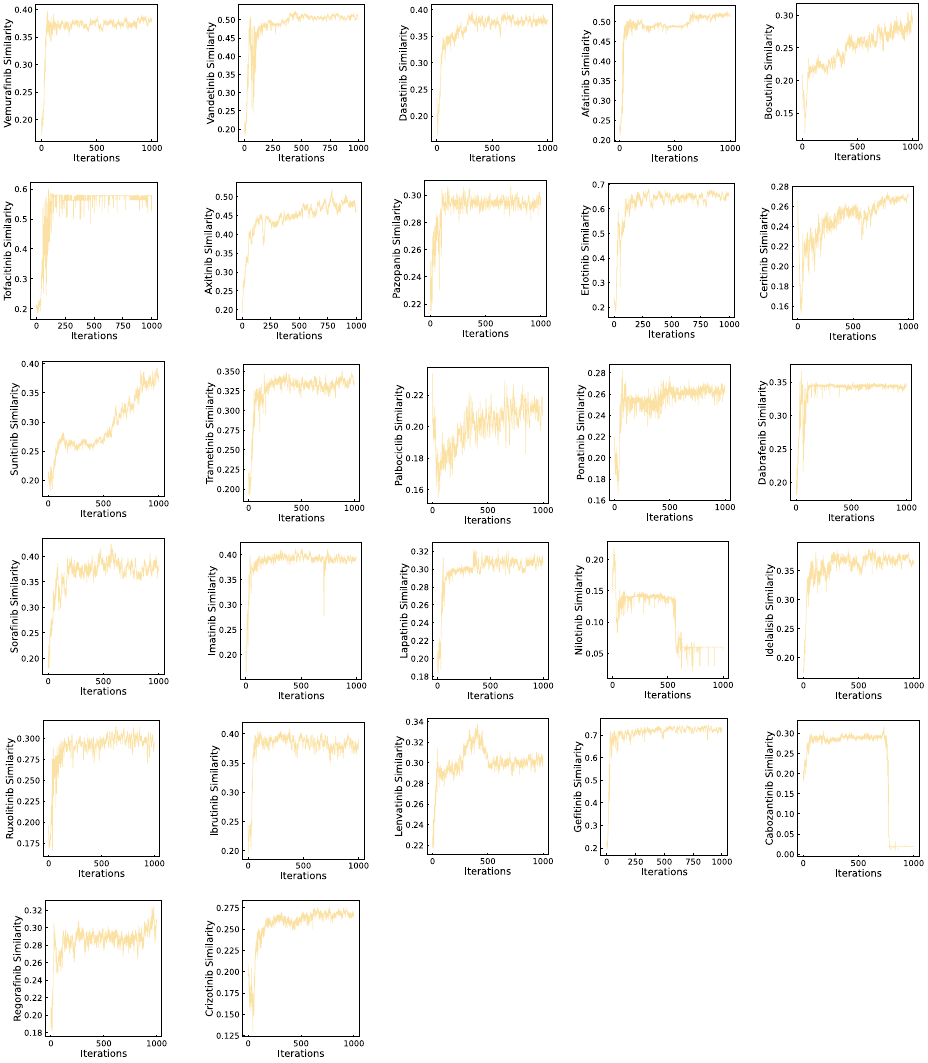}
  \caption{\textbf{Tanimoto similarity score computed between generated candidates and FDA approved TKI molecules, as a function of training iteration}}
  \label{fig:SI_REINVENT_scores}
\end{figure}

Figure \ref{fig:SI_REINVENT_scores} shows the similarity to query structure as a function of training iterations, for each of the 27 held-out FDA-approved TKI molecules. In most cases, we observed an increasing trend in the reward. There were however some instances (ex. Nilotinib and Cabozantinib) where training was unstable and did not converge. Reinforcement Learning algorithms are often highly sensitive to hyperparameters, so it is possible that these cases might require further tuning.

\subsection{OPD}
\label{SI:REINVENT_OPD}
\begin{figure}[H]
  \centering
  \includegraphics[width=0.9\textwidth]{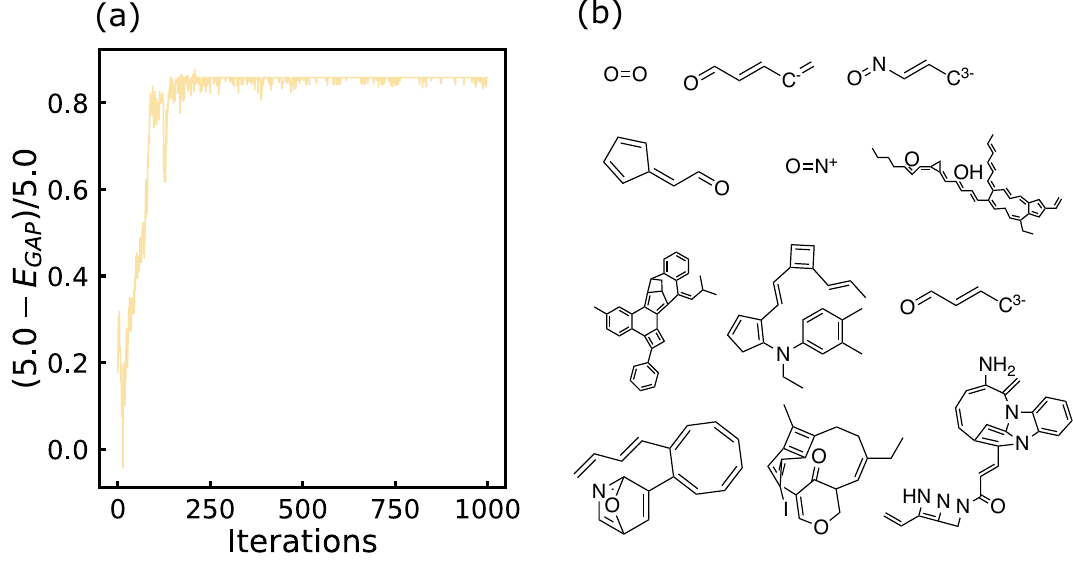}
  \caption{\textbf{Results of REINVENT+SELFIES on OPD dataset.} a) Reward score as a function of training iterations. b) Molecules sampled during later stages of training. }
  \label{fig:SI_reinvent_opd_mols}
\end{figure}

\begin{figure}[H]
  \centering
  \includegraphics[width=0.9\textwidth]{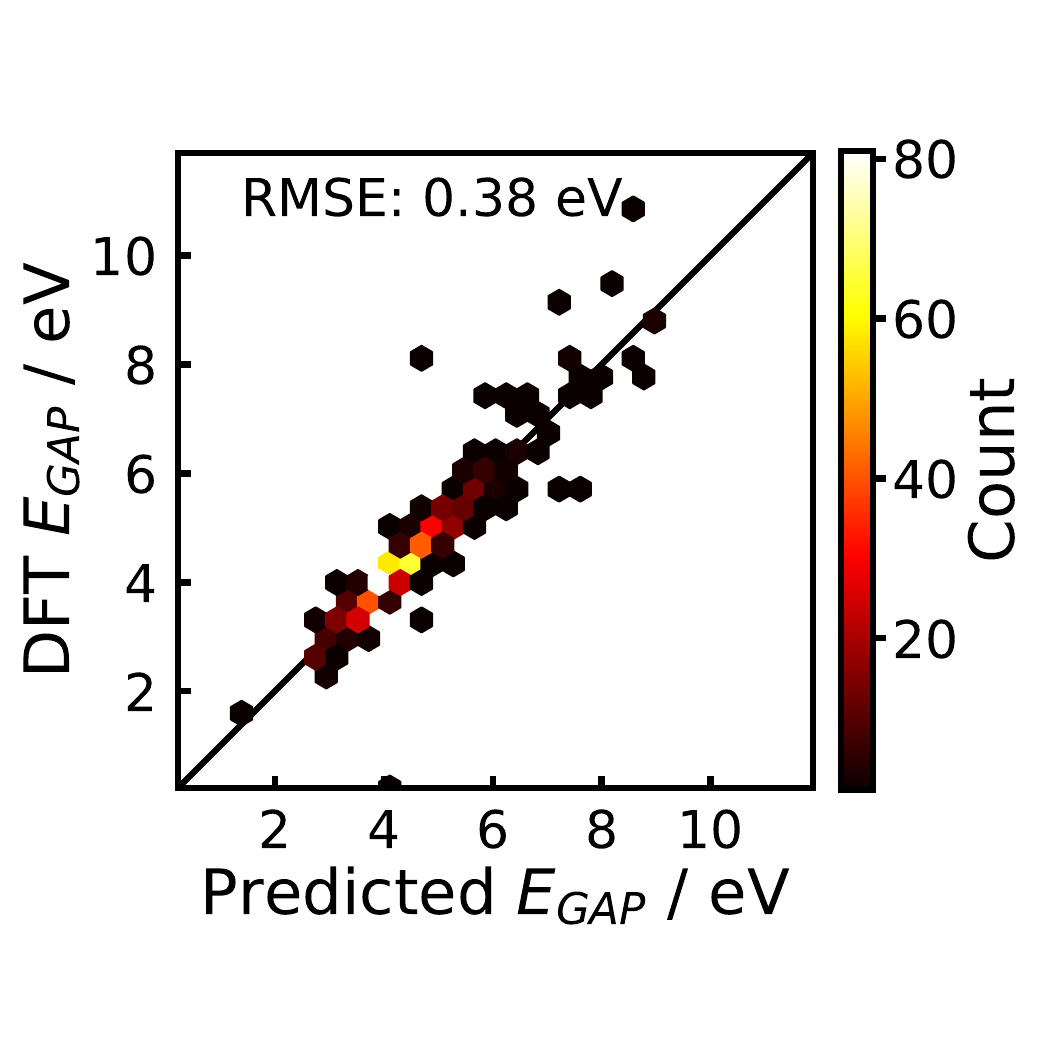}
  \caption{\textbf{Comparison between DFT-calculated optical gaps and chemprop-predicted optical gaps, as calculated on the test set. } RMSE on test set was 0.38 eV. }
  \label{fig:SI_chemprop_hexbin}
\end{figure}

Unlike the TKI dataset case where we had access to the oracle reward, training on the OPD dataset required a proxy neural network reward estimator. Figure \ref{fig:SI_chemprop_hexbin} shows the test performance of the proxy reward predictor on DFT-calculated optical gaps. We observed that while the reward had an increasing trend during training of the agent (Figure \ref{fig:SI_reinvent_opd_mols}(a)), the sampled molecules (Figure \ref{fig:SI_reinvent_opd_mols}(b)) did not match the training data well structurally. We hypothesised that this behavior arose from agent identifying and targeting high-uncertainty regions of the property predictor. To investigate this, we also attempted running agent training with a new reward that penalized high uncertainty as estimated by ensemble variance on the property predictor. To achieve this, the reward was modified to include a multiplicative masking term that evaluated whether the ensemble uncertainty was smaller than the 99th percentile of training data uncertainties. Hence molecules for which the property predictor was more uncertain than 99\% of the training data would have a reward of zero. We were however unable to achieve model convergence with this modified reward function, i.e., rewards did not display an increasing trend. This was because a majority of molecules generated during training were high-uncertainty points and resulted in a reward of zero. This resulted in the agent having access to very sparse information since poor candidates were sampled at a much higher fraction than good ones. It is also possible that the ensemble uncertainty was not an accurate estimator of model confidence at points that are highly Out of Distribution (OOD), as was observed by Scalia et al. in their work. \cite{scalia2020evaluating}

\section{JTVAE training}
\label{SI_JTVAE}
Since DFT calculations were expensive to perform on the entire patent-mined OPD set, we only labeled a subset of 5568 molecules out of a total of 112436 molecules. To effectively use labeled and unlabeled data during JTVAE training, we utilized all molecules for encoder and decoder training, but only utilized the labeled subset while training the property predictor. The training of encoder, decoder and property predictor were all performed jointly with a multitask loss function. In addition, the property predictor training was performed on 5 different properties: HOMO, LUMO, optical gap, Synthetic Complexity Score (SCScore) \cite{coley2018scscore}, and molecular mass. This was done for two purposes: 1)To aid with latent space regularization, and 2) Multiple tasks could potentially have shared information and thus compound the amount of effective training data seen by the model.

\section{FDA approved TKI candidates}
Table \ref{FDA_TKI} lists the names and SMILES representations of the 27 FDA-approved TKI molecules that were held out during all TKI experiments carried out in this paper. 
\begin{table}[htbp]
\resizebox{\textwidth}{!}{%
\begin{tabular}{ |c|c| } 
 \hline
 Name & SMILES \\ 
 \hline 
 Afatinib & CN(C)CC=CC(=O)NC1=C(C=C2C(=C1)C(=NC=N2)NC3=CC(=C(C=C3)F)Cl)OC4CCOC4 \\
 Ibrutinib & C=CC(=O)N1CCC[C@H](C1)N2C3=NC=NC(=C3C(=N2)C4=CC=C(C=C4)OC5=CC=CC=C5)N \\
 Pazopanib & CC1=C(C=C(C=C1)NC2=NC=CC(=N2)N(C)C3=CC4=NN(C(=C4C=C3)C)C)S(=O)(=O)N \\
 Axitinib & CNC(=O)C1=CC=CC=C1SC2=CC3=C(C=C2)C(=NN3)/C=C/C4=CC=CC=N4 \\
 Idelalisib & CC[C@@H](C1=NC2=C(C(=CC=C2)F)C(=O)N1C3=CC=CC=C3)NC4=NC=NC5=C4NC=N5 \\
 Ponatinib & CC1=C(C=C(C=C1)C(=O)NC2=CC(=C(C=C2)CN3CCN(CC3)C)C(F)(F)F)C\#CC4=CN=C5N4N=CC=C5 \\
 Bosutinib & CN1CCN(CC1)CCCOC2=C(C=C3C(=C2)N=CC(=C3NC4=CC(=C(C=C4Cl)Cl)OC)C\#N)OC \\
 Imatinib & CC1=C(C=C(C=C1)NC(=O)C2=CC=C(C=C2)CN3CCN(CC3)C)NC4=NC=CC(=N4)C5=CN=CC=C5 \\
 Regorafinib & CNC(=O)C1=NC=CC(=C1)OC2=CC(=C(C=C2)NC(=O)NC3=CC(=C(C=C3)Cl)C(F)(F)F)F \\
 Cabozantinib & COC1=CC2=C(C=CN=C2C=C1OC)OC3=CC=C(C=C3)NC(=O)C4(CC4)C(=O)NC5=CC=C(C=C5)F \\
 Lapatinib & CS(=O)(=O)CCNCC1=CC=C(O1)C2=CC3=C(C=C2)N=CN=C3NC4=CC(=C(C=C4)OCC5=CC(=CC=C5)F)Cl \\
 Ruxolitinib & C1CCC(C1)[C@@H](CC\#N)N2C=C(C=N2)C3=C4C=CNC4=NC=N3 \\
 Ceritinib & CC1=CC(=C(C=C1C2CCNCC2)OC(C)C)NC3=NC=C(C(=N3)NC4=CC=CC=C4S(=O)(=O)C(C)C)Cl \\
 Sorafinib & CNC(=O)C1=NC=CC(=C1)OC2=CC=C(C=C2)NC(=O)NC3=CC(=C(C=C3)Cl)C(F)(F)F \\
 Crizotinib & C[C@H](C1=C(C=CC(=C1Cl)F)Cl)OC2=C(N=CC(=C2)C3=CN(N=C3)C4CCNCC4)N \\
 Sunitinib & CCN(CC)CCNC(=O)C1=C(NC(=C1C)/C=C$\backslash$2/C3=C(C=CC(=C)F)NC2=O)C \\
 Dabrafenib & CC(C)(C)C1=NC(=C(S1)C2=NC(=NC=C2)N)C3=C(C(=CC=C3)NS(=O) (=O)C4=C(C=CC=C4F)F)F \\
 Tofacitinib & C[C@@H]1CCN(C[C@@H]1N(C)C2=NC=NC3=C2C=CN3)C(=O)CC\#N \\
 Dasatinib & CC1=C(C(=CC=C1)Cl)NC(=O)C2=CN=C(S2)NC3=CC(=NC(=N3)C)N4CCN(CC4)CCO \\
 Lenvatinib & COC1=CC2=NC=CC(=C2C=C1C(=O)N)OC3=CC(=C(C=C3)NC(=O)NC4CC4)Cl \\
 Trametinib & CC1=C2C(=C(N(C1=O)C)NC3=C(C=C(C=C3)I)F)C(=O)N(C(=O)N2C4=CC=CC(=C4)NC(=O)C)C5CC5 \\
 Erlotinib & COCCOC1=C(C=C2C(=C1)C(=NC=N2)NC3=CC=CC(=C3)C\#C)OCCOC \\
 Nilotinib & CC1=C(C=C(C=C1)C(=O)NC2=CC(=CC(=C2)C(F)(F)F)N3C=C(N=C3)C)NC4=NC=CC(=N4)C5=CN=CC=C5 \\
 Vandetinib & CN1CCC(CC1)COC2=C(C=C3C(=C2)N=CN=C3NC4=C(C=C(C=C4)Br)F)OC \\
 Gefitinib & COC1=C(C=C2C(=C1)N=CN=C2NC3=CC(=C(C=C3)F)Cl)OCCCN4CCOCC4 \\
 Palbociclib & CC1=C(C(=O)N(C2=NC(=NC=C12)NC3=NC=C(C=C3)N4CCNCC4)C5CCCC5)C(=O)C \\
 Vemurafinib & CCCS(=O)(=O)NC1=C(C(=C(C=C1)F)C(=O)C2=CNC3=C2C=C(C=N3)C4=CC=C(C=C4)Cl)F \\
 \hline
\end{tabular}}
\caption{\textbf{Names and SMILES strings of held-out FDA approved TKI molecules.}}
\label{FDA_TKI}
\end{table}

\section{Visualizing structural resemblance to training data}
\begin{figure}[H]
  \centering
  \includegraphics[width=0.9\textwidth]{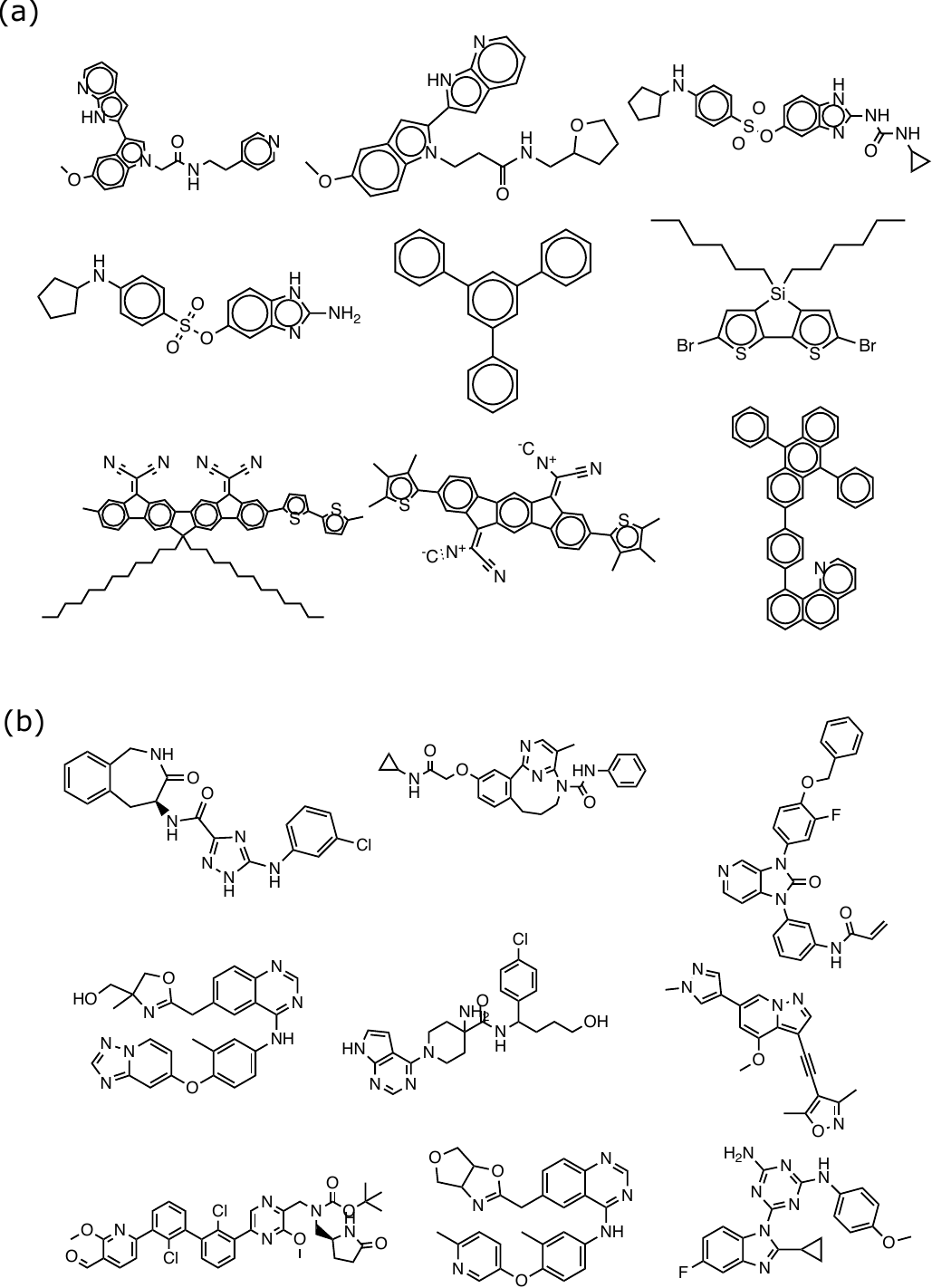}
  \caption{\textbf{Sample molecular structures obtained from random sampling of trained RNN+SELFIES model on a) OPD b) TKI dataset}}
  \label{fig:SI_RNN_SELFIES_mols}
\end{figure}

\begin{figure}[H]
  \centering
  \includegraphics[width=0.9\textwidth]{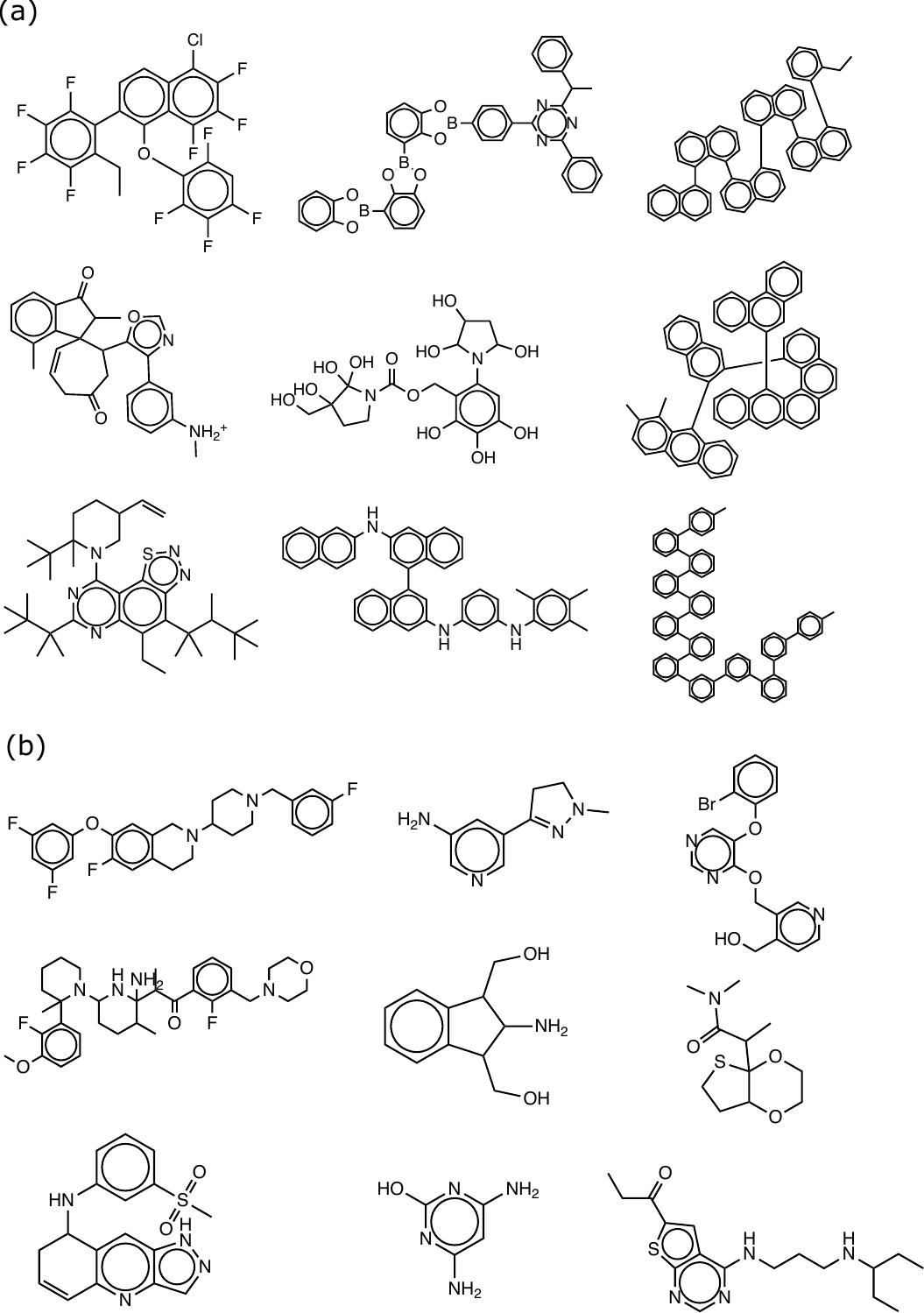}
  \caption{\textbf{Sample molecular structures obtained from random sampling of trained JTVAE model on a) OPD b) TKI dataset}}
  \label{fig:SI_JTVAE_mols}
\end{figure}

\end{document}